\documentclass[lettersize,journal]{IEEEtran}
\usepackage{amsmath,amsfonts}
\usepackage{algorithmic}
\usepackage{algorithm}
\usepackage{array}
\usepackage[caption=false,font=normalsize,labelfont=sf,textfont=sf]{subfig}
\usepackage{textcomp}
\usepackage{stfloats}
\usepackage{url}
\usepackage{verbatim}
\usepackage{graphicx}
\usepackage{cite}

\usepackage{soul}
\usepackage{multirow}
\usepackage{comment}

\usepackage{color,soul}

\usepackage{tabularray}

\usepackage{orcidlink}

\newcolumntype{C}[1]{>{\centering\arraybackslash}m{#1}}

\begin{document}
\title{Data streaming platform for crowd-sourced\\ vehicle dataset generation}


\author{Felipe Mogollon, Zaloa Fernandez\orcidlink{0000-0002-2201-4732}, Angel Martin\orcidlink{0000-0002-1213-6787}, Juan Diego Ortega, and Gorka Velez\orcidlink{0000-0002-8367-2413} 

\thanks{Felipe Mogollon, Zaloa Fernandez and Angel Martin are with the Department of Digital Media, Vicomtech Foundation, Basque Research and Technology Alliance (BRTA), San Sebastian, 20009 Spain (e-mail: \{fmogollon, zfernandez, amartin\}@vicomtech.org).

Juan Diego Ortega, and Gorka Velez are with the Department of Connected and Cooperative Automated Systems, Vicomtech Foundation, Basque Research and Technology Alliance (BRTA), San Sebastian, 20009 Spain (e-mail: \{jdortega, gvelez\}@vicomtech.org).

}
}

\markboth{This article has been accepted for publication in IEEE Transactions on Intelligent Vehicles. DOI: 10.1109/TIV.2024.3486926}
{Mogollon \MakeLowercase{\textit{et al.}}: Data streaming platform for crowd-sourced vehicle dataset generation}



\maketitle
\IEEEoverridecommandlockouts
\IEEEpubid{\begin{minipage}{\textwidth}\ \\\\\\\\\\[12pt]
\copyright 2024 IEEE.  Personal use of this material is permitted.  Permission from IEEE must be obtained for all other uses, in any current or future media, including reprinting/republishing this material for advertising or promotional purposes, creating new collective works, for resale or redistribution to servers or lists, or reuse of any copyrighted component of this work in other works.
 \end{minipage}}

\begin{abstract}
Vehicles are sophisticated machines equipped with sensors that provide real-time data for onboard driving assistance systems. 
Due to the wide variety of traffic, road, and weather conditions, continuous system enhancements are essential. Connectivity allows vehicles to transmit previously unknown data, expanding datasets and accelerating the development of new data models. This enables faster identification and integration of novel data, improving system reliability and reducing time to market. Data Spaces aim to create a data-driven, interconnected, and innovative data economy, where edge and cloud infrastructures support a virtualised IoT platform that connects data sources and development servers. This paper proposes an edge-cloud data platform to connect car data producers with multiple and heterogeneous services, addressing key challenges in Data Spaces, such as data sovereignty, governance, interoperability, and privacy. The paper also evaluates the data platform's performance limits for text, image, and video data workloads, examines the impact of connectivity technologies, and assesses latencies. The results show that latencies drop to 33ms with 5G connectivity when pipelining data to consuming applications hosted at the edge, compared to around 77ms when crossing both edge and cloud processing infrastructures. The results offer guidance on the necessary processing assets to avoid bottlenecks in car data platforms.
\end{abstract}


\section{Introduction} 
\IEEEPARstart{A}{dvanced} Driver Assistance Systems (ADAS) and Autonomous Driving (AD) functions must be validated through specific test procedures that assess their responses under various safety-related scenarios. The integration of perception capabilities in advanced functions has transformed car makers' validation approaches, as these systems often employ non-deterministic technologies like Machine Learning (ML) \cite{tits2024}. Despite the advanced perception and modelling capabilities of ML-based technologies, they must be rigorously tested across all possible driving conditions in which they are expected to operate. Within the classical “V” model validation approach \cite{ISO26262}, the training dataset is closest to system requirements, and the validation dataset is closest to the testing plan. Therefore, safety claims can only be made if the training and testing datasets comprehensively cover all safety-relevant scenarios \cite{Koopman2017}.

The necessity for extensive training and test datasets has driven the automotive industry to develop data collection and management methods that incorporate instrumented vehicles with ground-truth sensors and recording capabilities, large computation clusters, and Big Data solutions for data transmission, archiving, curation, analysis, and visualisation \cite{tits20242}. Traditional methods involve large-scale recording campaigns where instrumented vehicles gather vast amounts of multi-sensor data to capture various scenarios of interest. This data is then uploaded, filtered, labelled, and archived.

Raw sensor data streams generate enormous amounts of data that require labelling and processing. Although semi-automatic methods exist, current tagging processes are still predominantly manual, necessitating human validation of the training and test datasets \cite{Sager2021, Day2023}. Moreover, progress in ML models could be accelerated if data-gathering campaigns extend beyond instrumented vehicles to include regular cars \cite{tits20243}. Here, 5G and edge Computing could realize the vision of crowd-sourced data collection for ML model retraining and refinement. 5G offers enhanced data throughput capabilities, while edge computing provides distributed processing to efficiently identify sensor data related to edge cases or unaddressed scenarios, thereby reducing the amount of data stored and processed.

Data Spaces in the automotive sector will bridge a variety of use cases and stakeholders \cite{rettore2019}. For example, live training loops of new services in the research and development (R\&D) stage for Original Equipment Manufacturers (OEMs) or Tier-X suppliers, networked parking with dynamic prediction of departure time, optimal route or free parking spots for mobility operators, and driving safety awareness for municipalities and public authorities. 
 
Based on the principles of Data Spaces, this paper proposes a versatile data streaming platform suitable for a wide spectrum of car services and stakeholders under a business-to-business model. This platform enables the collection of relevant data from connected cars, which can be used to generate datasets for retraining and testing ML models. It also allows data to be filtered and prepared closer to the source, resulting in more efficient cloud resource usage and eliminating the need for dedicated vehicle fleets. The present work aims to study the feasibility of this novel concept, focusing on data challenges and specifically analysing the effects of different workloads in terms of concurrency, data complexity, connectivity technology, and cloud or edge processing, in terms of latency and scalability. This paper focuses on the IoT platform virtualised and operated on top of edge and cloud infrastructures, and it includes several novel contributions:
\begin{itemize}
    \item Translation of Data Spaces Principles: We adapted the principles of Data Spaces into specific data features that informed the design of a car data platform. This platform addresses the unique requirements and characteristics of data generated and consumed by intelligent vehicles and associated services.
    \item Edge-Cloud Architecture Mapping: We mapped these data features onto an integrated edge and cloud architecture to enable more efficient and scalable data processing. This approach supports the rapid delivery of data to Machine Learning Operations (MLOps)-based control and decision-making applications, as well as to Cooperative, Connected, and Automated Mobility (CCAM) systems that demand low-latency responses.
    \item Infrastructure Analysis for Low-Latency Processing: We analysed communication latency and computing capacity to determine the optimal sizing of the car data infrastructure. This ensures the system can effectively manage data throughput with minimal latency, using varying data standards, volumes, connectivity technologies, and the deployment of real-time data pipelines either at the edge or in the cloud.
\end{itemize}

The rest of the paper is organized as follows. Section \ref{sota} provides an overview of pipelines for automating and simplifying ML workflows and deployments, as well as IoT architectures for vehicular communications and CCAM applications. Section \ref{solution} explains the proposed data platform based on Data Spaces principles, which aims to increase processing capacity and reduce latency while fostering a car data marketplace, and describes how the MEC and cloud setups support the developed platform. Section \ref{eval} outlines the methodology and frameworks used to evaluate performance, capacity, and resulting latency of the IoT platform, along with the obtained results. Finally, Section \ref{conclusions} highlights the key results.

\section{Related work}
\label{sota}
\subsection{MLOps for CCAM}
MLOps is a critical practice that brings together the principles of Development Operations (DevOps) and data science. It focuses on streamlining the end-to-end lifecycle of ML models, from development to deployment and ongoing maintenance \cite{John2021}. One of the main benefits of MLOps is that it bridges the gap between ML models and production systems, ensuring a seamless deployment process that enhances efficiency and reduces time-to-market \cite{Makinen2021,Kreuzberger2023}.

The automotive industry is undergoing a transformative shift, fuelled by advancements in machine learning and connectivity. As vehicles become more intelligent and interconnected, the need for robust and efficient MLOps practices becomes paramount. The CCAM industry increasingly relies on ML models for predictive tasks \cite{Muhammad2021}. These models enhance safety, optimise vehicle performance, and provide personalised experiences for drivers and passengers. Within the development of Autonomous Vehicles (AV), MLOps plays a crucial role, as ML algorithms drive perception, decision-making, and control. Ensuring the reliability and safety of these models demands robust deployment and monitoring practices. By automating deployment pipelines, MLOps reduces downtime and optimises resource utilisation, leading to cost savings and operational effectiveness.

The deployment of ML models into production systems is a complex process, often leading to a gap between the development of the model and its use in real-world applications. MLOps bridges this gap by providing a set of practices and tools that facilitate the seamless deployment, monitoring, and maintenance of ML models \cite{Bustamante2023}. This is particularly crucial in the automotive sector, where ML models must interact with various sensors and systems in real-time, often under stringent safety and performance requirements \cite{Shi2021, Xiao2023}.

The challenges of deploying ML models in connected vehicles are multifaceted. Real-time data processing, limited computational resources, and the need for safety-critical applications pose significant hurdles. MLOps provides a framework to tackle these challenges, ensuring that ML models are robust, efficient, and safe for use in vehicles. It does so by facilitating continuous integration, continuous delivery, automated testing, and robust monitoring of ML models \cite{Kreuzberger2023}.

High-quality training data is paramount for the success of ML models. In the automotive sector, this data can come from diverse sources, including onboard sensors, connected infrastructure, and user inputs \cite{Hussain2019}. MLOps emphasises the importance of data collection, labelling, and preprocessing techniques specific to automotive use cases~\cite{Chen2022,Lee2023}. This ensures that the training data is relevant, accurate, and ready for use in model development.

In addition, data pipelines play a crucial role in automating the processes of data collection, preprocessing, and model training. They ensure consistency and efficiency in the ML workflow. Similarly, version control is essential for tracking changes to ML models, facilitating collaboration among team members, and enabling rollback in case of issues. MLOps provides the necessary tools and practices for managing data pipelines and version control, thereby enhancing the efficiency and reliability of ML workflows.

A SafeOps concept of "continuous safety" based on the DevOps approach is proposed in \cite{Fayollas2020}. In \cite{Rolf2022}, the authors propose that real-world field feedback for an initially safe deployment should support a DevOps-style continuous learning approach to lifecycle safety.

\subsection{IoT platforms}
Internet of Things (IoT) platforms deliver captured/sensed data to advanced vertical services for processing. These platforms connect sensors and devices to data services that perform data analytics and device management. The most important IoT platforms include AWS IoT, Azure IoT, Watson IoT, PTC ThingWorx, Google IoT, and Oracle IoT Cloud \cite{Barros2022}. They offer capabilities such as interoperability, security, privacy, developer support, data management, device management, and services management. These platforms enable organisations to focus on business insights (cloud analytics and custom business logic) instead of data management and device inventory.

Amazon AWS IoT is an IoT cloud platform designed to operate IoT devices and create value from the events they publish. It provides a REpresentational State Transfer (REST) Application Programming Interface (API) for configuration and operation. Additionally, the AWS IoT message broker supports MQTT and LoRaWAN protocols, enabling interaction between devices and applications through the publish and subscribe model. IBM Watson IoT Platform is also a cloud-based set of services that gathers data from devices by sending it to the cloud using the MQTT protocol. It includes functions for registering and managing IoT devices, securely sending data, and visualising it on integrated dashboards. Data is available in real time and historically through REST APIs, allowing applications to consume, process, and present it according to their processing mode. Microsoft Azure IoT supports common IoT communication protocols such as HTTP, MQTT, and AMQP, and multiple software stacks including Azure RTOS, FreeRTOS, and BareMetal. PTC ThingWorx Industrial Platform is a comprehensive device integration cloud platform based on industry-compliant standards. This is achieved using the ThingWorx connection server based on the proprietary  WebSocket protocol (AlwaysOn Protocol) and MQTT. Google Cloud IoT Core establishes connections using a bi-directional protocol bridge with MQTT and HTTP endpoints. Its main functions are registration, authentication, and authorisation, enabling third-party services to interact with any registered device. Due to Google's interoperability, implementing business logic at the edge and performing data analytics is feasible.

Essentially, all these platforms provide inventory and connection to sensors and devices, offer sophisticated or simple data analytics, and natively integrate with their cloud services to create control loops similar to those implemented by Supervisory Control and Data Acquisition (SCADA) systems \cite{Sverko2022}. However, they focus on vertical services that own the data sources \cite{Joy2017}. The future of the automotive data economy will bring a new variety of stakeholders and actors who will collaborate and exploit shared data to create new value in the value chain \cite{Marosi2018}. This brings many implications and new requirements.

There are some open-source projects aimed at filling the gaps. Eclipse Kuksa aims to provide shared building blocks for Software Defined Vehicles that can be shared across the industry \cite{kuksa2020}. The Eclipse Kuksa ecosystem comprises three main platforms: the (1) in-vehicle platform, (2) cloud platform, and (3) an app IDE. The cloud platform uses a microservices architecture to adapt to demand in real-time and offer telemetry services. Kuksa indexes and accesses vehicle data using the W3C Vehicle Information Service Specification (VISS) v2 protocol based on WebSockets \cite{viss2022}.

\section{Proposed solution}
\label{solution}
IoT technologies are ideal for communicating data from multiple sensors to control and management systems. They play a crucial role in car manufacturing, making them familiar to OEMs and Tier-1 suppliers. Accordingly, the use of IoT messaging solutions widely employed in industrial processes is practical and convenient for data-centric applications and services in the CCAM sector.

On top of a distributed architecture, the proposed data platform provides a data-centric Internet of Things (IoT) messaging platform for CCAM services and application developers. The data platform essentially connects data producers and data consumers. Unlike data lakes, which are widely used for offline and batch processing, it delivers data produced in real time to data consumers. Cars can serve as either producers or consumers, depending on whether they are publishing sensor data or receiving event notifications from CCAM applications deployed in the cloud or MEC.

The platform can grant or revoke access to data, as consumers are not directly connected to data producers (sources). This ensures total decoupling between data producers and data consumers. Control over who can access data is managed based on licences. To encourage data producers to share data, anonymisation techniques can be applied.

Therefore, the data platform adopts a hierarchical architecture, utilising the cloud layer to centralise data production and consumption while leveraging the MEC layer to capitalise on the distributed edge infrastructure for performing data anonymisation and sampling as close to the originating sensor or device as possible. Additionally, the platform offers key features for data consumers including geo-pinned data filters for querying, the ability to deploy low latency services at specific locations, and data quality scores. The data quality assessment is based on the European Telecommunications Standards Institute (ETSI) standard TS 103 759 V2.1.1 (2023-01) \cite{ETSIquality2023}. This approach incorporates trust mechanisms that detect contradictions in data through information redundancy, such as location proximity, in contrast to cryptographic solutions for trustworthiness \cite{liu2023}, or the application of ML-assisted methods for detecting sensor inconsistencies \cite{guo2020}.

Finally, it enhances platform operation efficiency, reducing operational expenditure as consumers scale up, through cost-performance schemas on data scalability, reusing the required processing for common pricing plans and data queries. The implemented data platform provides mechanisms, data, and tools to support a pay-as-you-go approach for pricing. Consequently, the monetisation vectors considered are licences, the volumes of data consumed by each user, and the computing assets allocated by the user to preprocess data before consumption.

\begin{figure}[!t]
    \centering
    \includegraphics[width=1.0\linewidth]{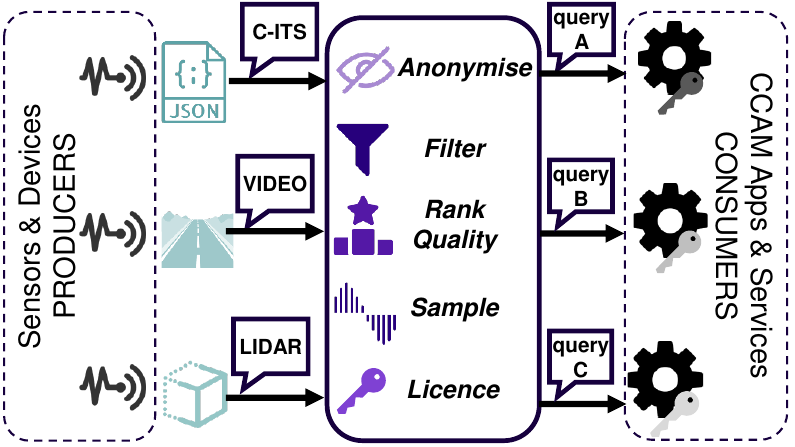}
    \caption{Features of the proposed data platform for developing CCAM applications and services.}
    \label{fig:features}
\end{figure}

All these features are illustrated in Figure \ref{fig:features}. In this setup, data producers push data samples to specific IoT topics (represented as speech bubbles) based on the data type—such as Cooperative Intelligent Transport Systems (C-ITS), video, or point clouds from Light Detection and Ranging (LIDAR) sensors—and select the appropriate licence option to grant access. The data platform then forwards the data to the relevant IoT topics for consumers, with data samples being anonymised when applicable. Consumers can use database-like queries to select one or multiple data types from specific areas. They can also choose only high-quality samples, according to platform statistics and derived reliability metrics. When sampling is requested according to a Service Level Agreement (SLA), the platform adjusts the subsampling level accordingly. It is important to note that consumers must have the correct licence to query eligible data flows (depicted as grey-scale keys).

\begin{figure}[!t]
    \centering
    \includegraphics[width=1.0\linewidth]{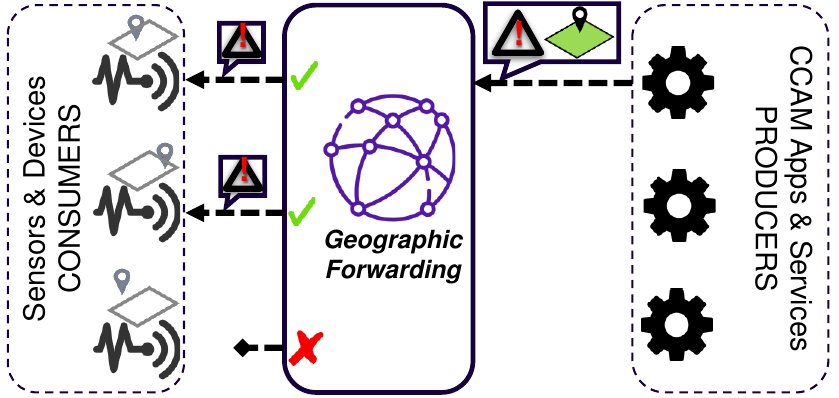}
    \caption{CCAM applications and services casting events to sensors and devices in an area.}
    \label{fig:casting}
\end{figure}

In Figure \ref{fig:casting}, the definition of Regions of Interest (ROIs) is crucial for directing messages from CCAM applications and services to only those sensors and devices that may find this information relevant within a specific geographic area. This data flow also relies on structured IoT topics (represented as speech bubbles).

\subsection{CCAM-driven design}
Future car data platforms will aim to connect automotive data spaces initiatives such as Catena-X \cite{catenax}, where \textit{data sovereignty, governance, reuse, interoperability, transparency, and privacy} must be design principles, while also \textit{supporting data monetization models}. Specifically, our proposed data solution specialises in feeding CCAM applications with real-time data to provide live information or trigger commands. This implies some simplifications compared to connecting data lakes but places the focus and pressure on low-latency streaming of data to meet CCAM demand expectations.

In a highly mobile and dense environment like connected cars, the Key Performance Indicators (KPIs) from 5G and beyond networks are crucial to avoid communication bottlenecks. Moreover, in the context of CCAM, the 5G architecture, including Multi-access Edge Computing (MEC), plays a prominent role in restricting the range of raw data to the vehicle's edge \cite{felipe2023}. This fosters data privacy by processing data as close as possible to where it is generated and transmitted. The data platform needs to manage privacy, as the onboard hardware is often sized to exclusively execute on-vehicle systems, and any external feature or system adds overhead. Furthermore, the edge computing architecture distributes the processing workload, eliminating the need for large computing infrastructures to handle all forwarded data. Instead, the edge infrastructure can be sized to process only a local environment.

To manage all the locations where the platform runs distributed data processing pipelines and to provide continuity of data as it crosses different edge infrastructures, the data platform must use the cloud to map and manage the distributed platform systems. This means decentralised processing and centralised data streaming.

To implement this approach, the data platform includes the features listed in Table \ref{tab:features} and classified around the CCAM-driven design categories.

\begin{table*}[!t]
    \caption{Features of the Data platform meeting data spaces' requirements.}
    \label{tab:features}
    \centering
    \renewcommand{\arraystretch}{1.4} 
    \begin{tabular}{|l|l|l|}
        \hline
        \textbf{Category} & \textbf{Feature} & \textbf{Description/Scope} \\
        \hline
        \multirow{5}{*}{\textbf{Interoperability}} & Topics & The platform receives and provides data at IoT topics according to the data type to enable data pipelines\\\cline{2-3}
         & IoT messaging & Lightweight and available for different architectures, Operating Systems and software stacks \\\cline{2-3}
         & Formats \& Metadata & Standard formats and metadata are used to foster harmonisation and setup configuration for processing\\\cline{2-3}
         & Two-ways & Event alarms and messages are cast to all the systems in a configured area\\\cline{2-3}
         & Open APIs & Open APIs for data register, ingest, browsing, querying and SLA setup to CCAM system developers \\
        \hline
        \multirow{3}{*}{\textbf{Reuse}} & On demand & Data producers only send data (incurring in overheads) when a consumer has performed a matching query\\\cline{2-3}
         & Concurrent queries & Data samples are processed once (at the edge) and multiplied (at the cloud) to consumers' IoT topics\\\cline{2-3}
         & Virtualized Pipelines & Pipelines specialised in a specific data type run on demand through containers at a specific edge\\
        \hline 
        \multirow{4}{*}{\textbf{Governance}} & Brokerage & Forwarding produced data samples to IoT topics satisfying data queries from data consumers\\\cline{2-3}
         & Geo-filtering & Attach geographic tags to allow the definition of regions of interest of data produced\\\cline{2-3}
         & Mobility handover & Enable tracking of individual data flows as they move  \\\cline{2-3}
         & Cloud-Edge & A cloud platform index edge infrastructures, query data flows, create IoT topics and instantiate pipelines\\
        \hline
        \multirow{2}{*}{\textbf{Sovereignty}} & No persistence & Data producers and owners can decide at any time to share registering sensors or to stop publishing \\\cline{2-3}
         & licence & Applying filters between producers and consumers to grant visibility of data licences \\
        \hline
        \multirow{2}{*}{\textbf{Transparency}} & Telemetry & Data flows, volumes, processing assets, requested regions and contracted SLAs are accounted \\\cline{2-3}
         & Quality scoring & Statistics for individual data flows enables queries for highly reliable and available data sources\\
        \hline
        \multirow{3}{*}{\textbf{Privacy}} & Anonymisation & Remove blacklisted attributes from text, faces and plates from images and video\\\cline{2-3}
         & Pseudoanonymisation & Assign uncorrelated identities to registered data flows to allow processing as time series \\\cline{2-3}
         & Protection & Access to IoT topics is protected with tokens to avoid promiscuous modes and account for all data access\\
        \hline
        \multirow{3}{*}{\textbf{Economy}} & SLA & Pipelines' containers run on computing assets depending on the data complexity and contracted capacity\\\cline{2-3}
         & Sampling & Subsampling levels to reduce billed items and accommodate throughput to CCAM processing capacity\\\cline{2-3}
         & Edge hosting & Low latency CCAM applications deployed in an edge controlling the access to locally queried topics\\
         \hline
    \end{tabular}
\end{table*}

\subsection{Edge pipeline}

A CCAM data platform needs to integrate different networks and infrastructures from various network operators and carriers to become a universal and comprehensive car data platform. However, MEC infrastructures are provided, managed, or operated by specific Mobile Network Operators (MNOs) and are accessible only to the UEs subscribed to their networks, serving services to them. This means that in a multi-public land mobile network (PLMN) setup, each MEC infrastructure is in an isolated domain, limited to the UEs subscribed to the MNO’s radio network. This isolation protects MEC infrastructures from security threats or attacks originating beyond the MNO’s subscribers or from the Internet. Additionally, the strict security policies and restrictions established by MNOs regarding external connections block any direct or fast connection between different MEC infrastructures, especially if they are operated by different MNOs.

Practically, this means that a User Equipment (UE) moving into the serving area of another MEC infrastructure would need to start a new session with the services provided, as there is no automatic session continuity or handover management from the network. 
When a user equipment (UE) is moving and streaming data across different Multi-access Edge Computing (MEC) infrastructures, it must manage the ongoing sessions. As the UE enters a new area, it is responsible for initiating a new session within the serving MEC infrastructure of MNO in that region.

To support this, a discovery mechanism in the cloud acts as the main entry point for any sensor or device to find the serving platform services. To this end, the onboarding of new edge platform instances is automated by a script deploying the entire software stack. The script for instantiating the data platform includes three stages, two static and sequential, and a final one dynamically executed when a new data consumer makes a request through the APIs provided in the cloud. These stages are the following ones:

\begin{enumerate}
    \item Deploy the base software stack, enabling the automated deployment of data platform modules at the MEC.
    \item Configure the new instance as part of the Data platform. This step primarily includes setting up connectivity to allow inbound and outbound connections between the cloud platform and the instantiated MEC platform.
    \item Create and destroy data pipelines on demand according to the CCAM applications and service requests forwarded from the cloud platform.
\end{enumerate}

The deployment of pipelines in the MEC is triggered by Data Consumers interested in processing and consuming a specific data type. The deployed pipelines performing anonymisation and data sampling are deployed as containers providing specialised data functions. If resources are available in the MEC serving the selected area, the data-type pipeline is deployed according to the contracted SLA. The container is downloaded from a registry, and the signatures of the containers are checked by the virtualisation service of the MEC data platform before deploying or instantiating them to run the requested data pipelines.

Alternatively, third parties can request the deployment of a CCAM application, composed of containers, in a specific MEC platform. These deployments enable swift information retrieval, significantly reducing latency. To ensure the security and integrity of the MEC environment, a stringent deployment procedure and runtime monitoring are followed. A quarantine procedure runs the low-latency CCAM application in a dedicated test environment with synthetic data. Here, the platform conducts comprehensive testing to validate the application’s correct behaviour, including verifying the proper utilisation of MEC resources and the constrained access to queried data flows. Once the application passes the tests, it is deployed on demand by the Third Party, using the same technical procedure used to deploy data pipelines in the MEC.

\subsection{Cloud services}

The cloud platform includes clusters, load balancers, and auto-scaling policies to efficiently manage the incoming demand. The main services deployed in the cloud for the data platform include:
\begin{itemize}
    \item Inventory of all edge infrastructures and data services to enable the data platform to connect the cloud and edge IoT topics, sensors, and devices to obtain the nearest data services.
    \item Management of licences to remove data flows from sources not covered by the consumer licence.
    \item Telemetry of computing assets, data volumes and licences for the individual IoT topics of each data consumer.
    \item Data catalogue enabling queries based on geographical areas, data types, licences, and qualities.
    \item Ordering the deployment or retirement of data pipelines according to the SLA contracted by the consumer, including sampling rates and a set of computing resources at the edge to process (and anonymise) data.
    \item IoT messaging merging data topics from different MEC infrastructures to satisfy data queries and multiplying the data in the different topics when different consumers target common data.
    \item Protecting access to IoT topics based on tokens.
\end{itemize}

Specifically, the system that directly translates data queries into workload is the IoT messaging. Thus, the scaling policies need to monitor the containers and resources required to avoid bottlenecks and ensure platform availability.

\section{Experimentation}
\label{eval}
This section details the infrastructures and tests carried out. It includes the evaluation criteria based on different metrics that are probed, assessed, and statistically analysed. The obtained results enable the outlining of the operational ranges, performance achieved, and estimated efficiency for a certain workload, thus providing reference points to forecast performance and required assets for operating the platform to meet the estimated demand for data production and consumption. Consequently, the results could guide other platforms handling automotive data regarding infrastructure sizing. To this end, we extensively expand on previous work \cite{felipe2023}, leveraging tools \cite{9826155}, procedures, and methodologies \cite{9303425} to conduct tests on the virtualised IoT platform \cite{9698101, 8789918}.

\subsection{Experimentation setup}

The experimentation comprises the deployment of the cloud and edge platforms and the instantiation of data producers and consumers to inject different workloads. The cloud platform consists of Amazon EKS (Elastic Kubernetes Service) with 2 nodes, each equipped with 2 CPUs and 8 GB of RAM, capable of automatically scaling resources vertically as needed. For the edge, a MEC infrastructure based on Kubernetes is connected to a 5G Stand-Alone (SA) Release 17 mobile network, hosting virtualised containers dynamically deployed when data consumers request connectivity to queried produced data. This infrastructure includes 36 CPUs and 128 GB of RAM.

\begin{figure}[!t]
    \centering
    \includegraphics[width=0.9\linewidth]{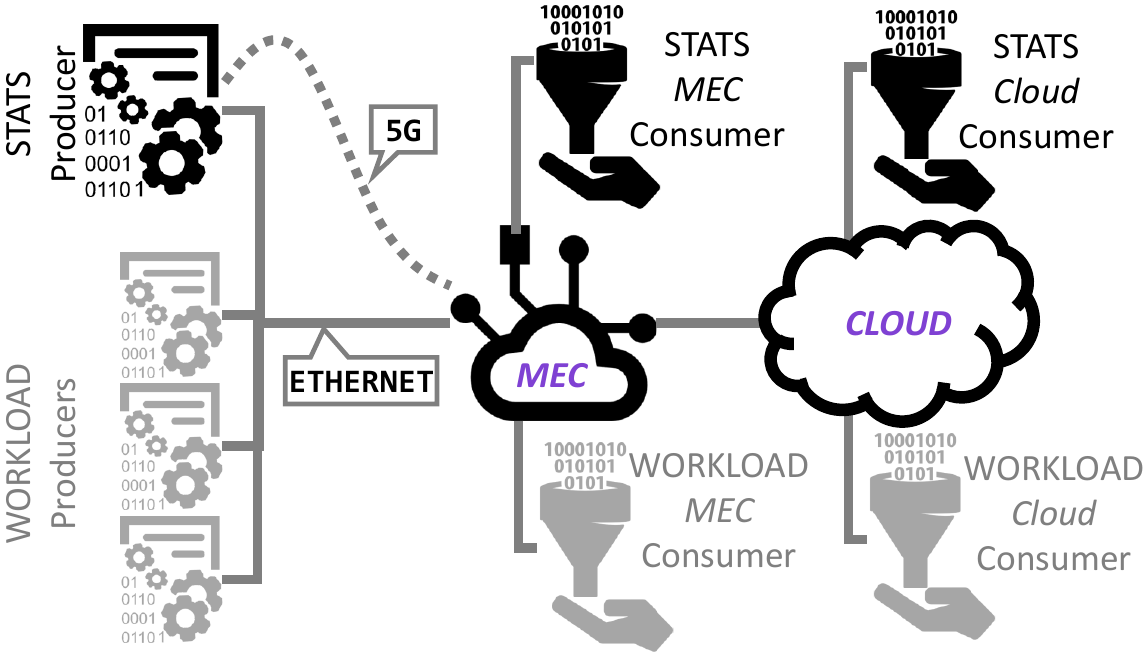}
    \caption{Experimentation setup including data producer and consumer and concurrent workload.}
    \label{fig:setup}
\end{figure}

To cover different networking possibilities, both wired and mobile connectivity are considered in all the tests, allowing data producers to send data to the platform via Ethernet or 5G New-Radio (NR). However, data consumers, representing data-centric applications, are mainly connected through Ethernet to either the cloud or the MEC platform. It is important to note that data producers can also be hosted in the MEC infrastructure, providing quicker access to data for latency-sensitive applications and narrower access to locally produced data. Figure \ref{fig:setup} provides an overview of the experimentation testbed.

The different data types considered include; Cooperative Intelligent Transport Systems (C-ITS) delivered in the EN 302 637-2 standard by the ETSI \cite{CITS}; images in Joint Photographic Experts Group (JPEG) format with a resolution of 1280x800 pixels; and video in H.264 format (ITU-T SG 16 \& ISO/IEC 14496-10) with a resolution of 1280x800 pixels, resulting in 2 Mbps for each produced data flow. The parameters for each data producer, along with the density of concurrent data producers contributing to the workload, are compiled in Table \ref{tab:scenarios}. This configuration results in a maximum throughput of 12.8 Mbps for C-ITS, 480 Mbps for JPEG, and 40 Mbps for H.264 data formats.

\begin{table}[!t]
    \caption{Data types, features and workloads employed in the tests.}
    \label{tab:scenarios}
    \centering
    \renewcommand{\arraystretch}{1.2} 
    \begin{tabular}{|l|c|c|r|r|r|}
        \hline
        \textbf{Data} & & \textbf{Producer} & \multicolumn{3}{c|}{\textbf{Number of Producers}} \\\cline{4-6}
        \textbf{Type} & \textbf{Format} & \textbf{Datarate} & \textbf{Low} & \textbf{Medium} & \textbf{High} \\
        \hline
        Message & C-ITS & 4 messages/s & 135 & 250 & 400 \\
        \hline
        Image & JPEG & 2 images/s & 25 & 50 & 70 \\
        \hline
        Video & H.264 & 10 frames/s & 6 & 10 & 20 \\
        \hline
    \end{tabular}
\end{table}

\subsection{Metrics}
The most relevant metrics for a business-oriented data platform streaming automotive data are related to two situations. First, at the beginning, when a request from a consumer is translated into the deployment of containers, setup of IoT services according to the query, and forwarding of data to the consumer pipeline. This period is referred to as \textit{Data Access Delay}. Second, once the data is being consumed, the time elapsed from the instant each data sample is produced to the instant it is consumed becomes critical for specific CCAM applications. This metric is known as \textit{Data Delivery Delay}. To accurately measure these timespans, timestamps added by the platform to each sample rely on a common Network Time Protocol (NTP) clock across the entire data platform.

Regarding the concurrency expectations for such a market-oriented data platform, scalability is a key parameter. The telemetry system captures the usage of virtualised assets, specifically \textit{CPU and RAM}. The objective is to assess the impact of different workloads and forecast specific demands. Since the testing setup is distributed across cloud and MEC infrastructures, assets from both architectures are monitored.

\subsection{Results}
The full set of results is compiled in Tables \ref{tab:resultsCloud} and \ref{tab:resultsMEC}. Table \ref{tab:resultsCloud} presents the performance and scalability results for different data types (refer to Table \ref{tab:features}), workloads (refer to Table \ref{tab:scenarios}), and connections used by the On-Board Unit (OBU) of Data Producers, utilising either 5G or Ethernet, with Data Consumers connected to the cloud data platform. In contrast, Table \ref{tab:resultsMEC} includes all the results where Data Consumers are hosted by the MEC platform to achieve low latency when processing local data.

\begin{figure}[!t]
    \centering
    \includegraphics[width=1\linewidth]{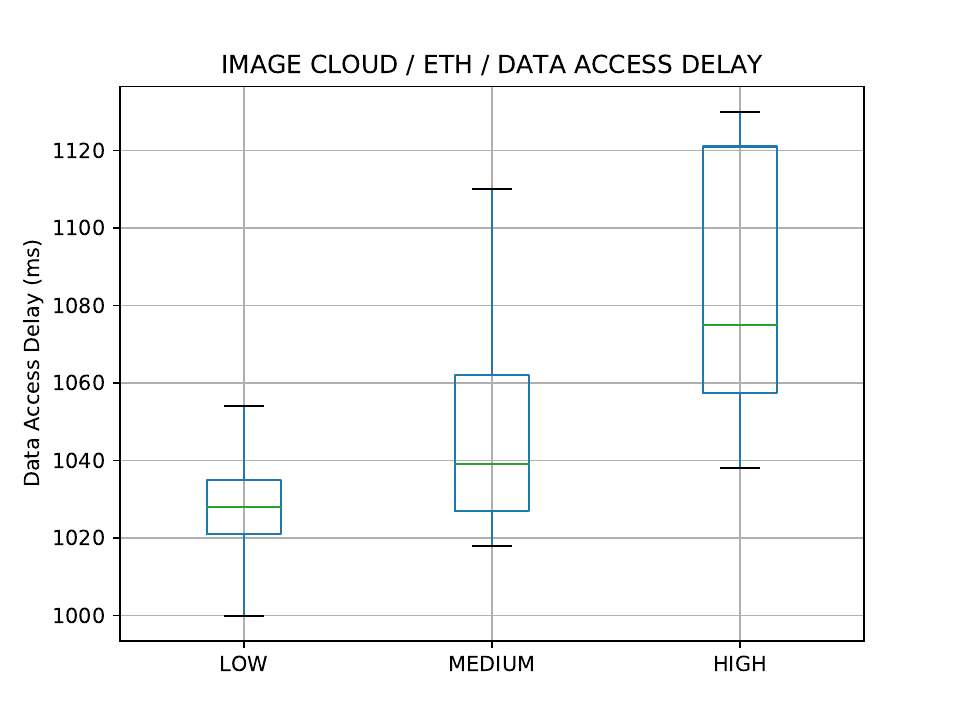}
    \caption{Data Access Delay for image data with producers connected by Ethernet and consumers accessing through the cloud.}
    \label{fig:imageAccessDelay}
\end{figure}

\begin{figure}[!t]
    \centering
    \includegraphics[width=1\linewidth]{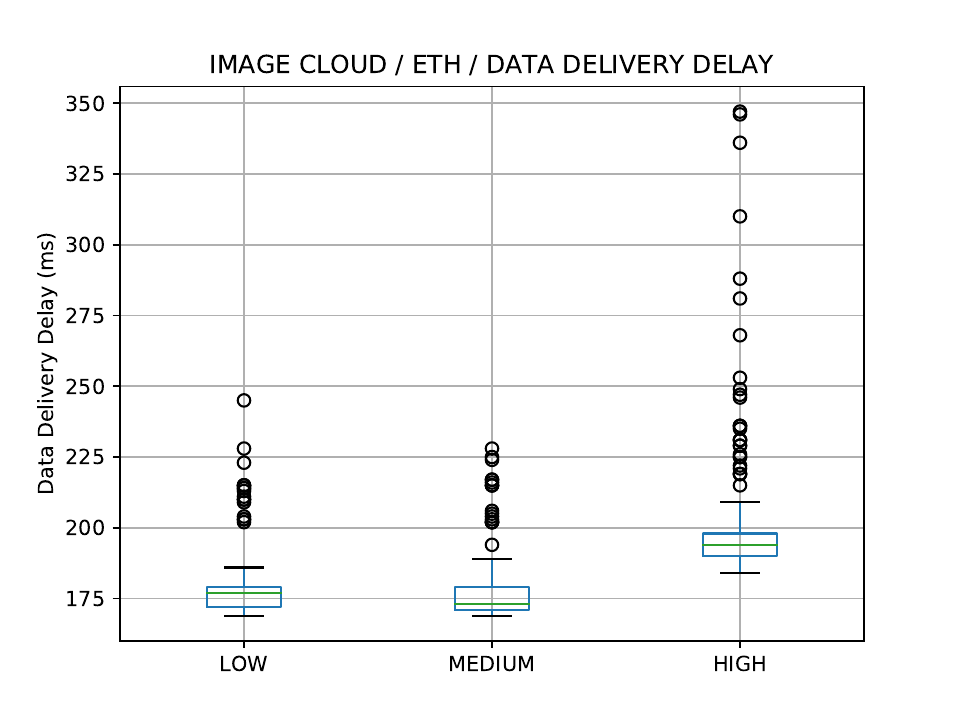}
    \caption{Data Delivery Delay for image data with producers connected by Ethernet and consumers accessing through the cloud.}
    \label{fig:imageDeliveryDelay}
\end{figure}

First, we assess if the applied workloads introduce enough dynamism to impact the performance or scalability metrics of the data platform. In this regard, the results depicted in Figures \ref{fig:imageAccessDelay} and \ref{fig:imageDeliveryDelay} clearly show that the workloads defined in Table \ref{tab:scenarios} result in significantly higher values and more outliers for \textit{Data Access Delay} and \textit{Data Delivery Delay}. Based on these results, projections on delay performance when applying such a platform to other data characteristics or densities can be made. Specifically, Figure \ref{fig:imageAccessDelay} illustrates how the defined workloads notably impact \textit{Data Access Delay}, which ranges from around 1 second to 1.1 seconds, indicating a non-negligible bootstrapping time for applications requiring quick and spontaneous access to data. This value includes the time required by the platform to instantiate virtualised data pipelines and connect data topics between the MEC and the cloud. In addition, it is evident that with a lower volume of data producers, the timing becomes more predictable and consistent.

Similar trends are observed with \textit{Data Delivery Delay}. Once the session is established and data streams are flowing, as depicted in Figure \ref{fig:imageDeliveryDelay}, the processing time added by the platform ranges from around 175 milliseconds to 200 milliseconds. This latency is sufficient for many applications, as it supports interactive applications requiring rapid communication. These values and behaviours also apply to video and C-ITS data. Similar patterns are observed for video and C-ITS data in terms of \textit{Data Access Delay} and \textit{Data Delivery Delay} ranges. It is important to underline that the volume and scattered outliers increase as the workload shifts from \textit{low} to \textit{high}.

\begin{figure}[!t]
    \centering
    \includegraphics[width=1\linewidth]{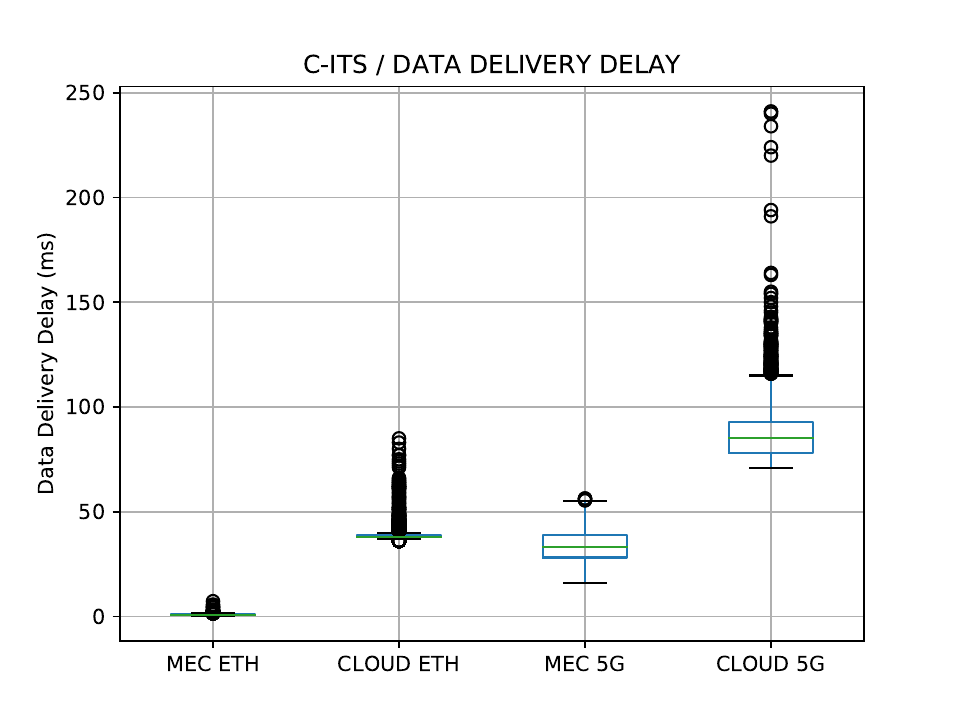}
    \caption{Data Delivery Delay for C-ITS data with producers using alternative network interfaces and consumers accessing through different infrastructures.}
    \label{fig:citsDeliveryDelay}
\end{figure}

Next, Figure \ref{fig:citsDeliveryDelay} highlights the significant role of the MEC in reducing latency when connecting Data Producers and Consumers for C-ITS data under \textit{high} workload conditions. The \textit{Data Delivery Delay} is reduced by approximately 40 milliseconds when Data Consumers are hosted in the MEC infrastructure. Additionally, performance is more variable, with more outliers, when connecting Data Consumers through the cloud platform. This variability could be crucial in deciding where to deploy a CCAM application. Moreover, Figure \ref{fig:citsDeliveryDelay} reveals the impact of 5G SA radio communications, as depicted in Figure \ref{fig:setup}, compared to Ethernet, which provides a faster network interface. The overhead of the cellular setup is under 50 milliseconds in both cases. Furthermore, in terms of reliability, derived from the packet loss, 5G achieves over 99.9788\% reliability, compared to the 100\% reliability obtained with Ethernet.

\begin{figure}[!t]
    \centering
    \includegraphics[width=1\linewidth]{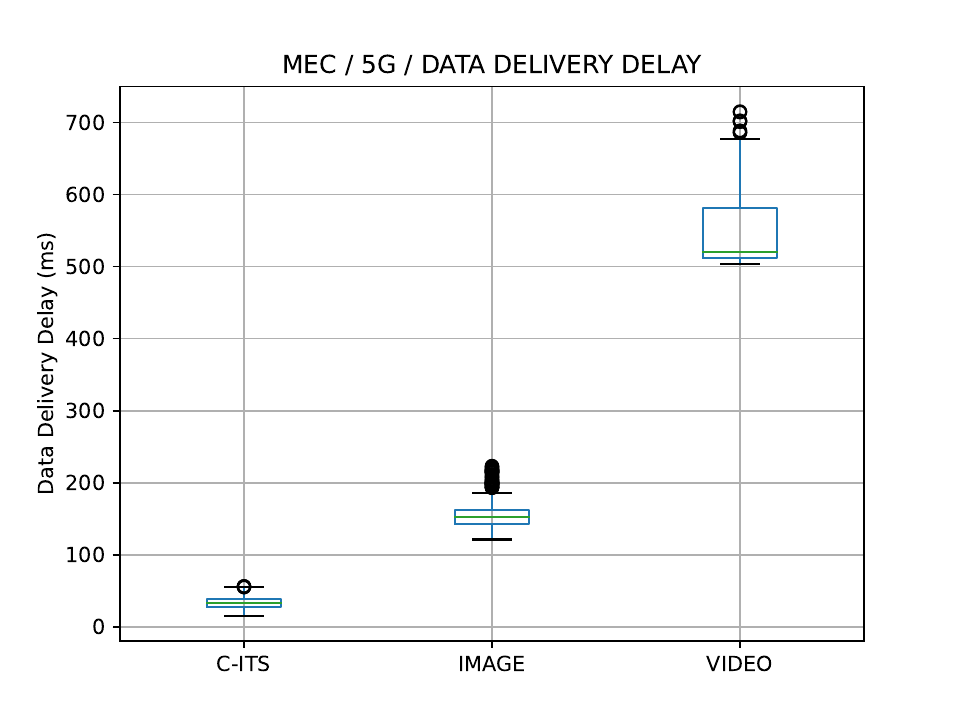}
    \caption{Data delivery delay for C-ITS, images and video data in the MEC for high workload of producers connected by 5G.}
    \label{fig:latencySummaryMEC5G}
\end{figure}

Figure \ref{fig:latencySummaryMEC5G} compares \textit{Data Delivery Delay} under \textit{high} workload for different data types when Data Producers are connected through 5G SA and Data Consumers are hosted in the MEC infrastructure. This setup results in a latency of under 35 milliseconds for C-ITS messages, 160 milliseconds for Image data types, and 550 milliseconds for Video streams. This latency should be compatible with low-latency CCAM applications running in the MEC infrastructure.

\begin{figure}[!t]
    \centering
    \includegraphics[width=1\linewidth]{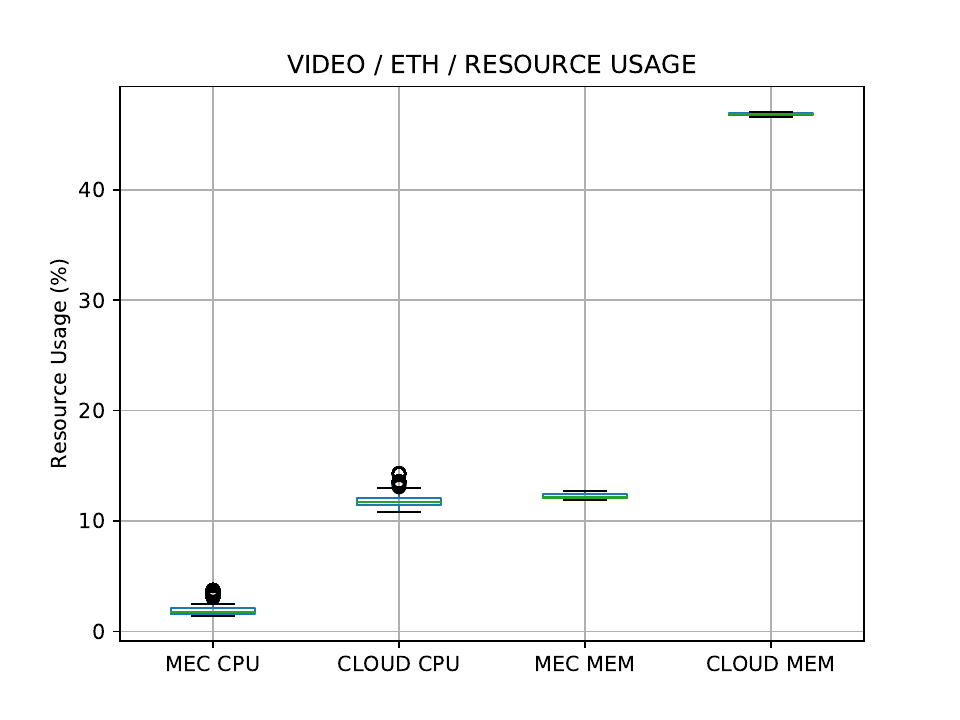}
    \caption{CPU and RAM consumption of the data platform systems in the MEC and the cloud for High workload of video producers connected by Ethernet.}
    \label{fig:videoCpuRAM}
\end{figure}

Finally, to determine whether the results are influenced by processing bottlenecks, telemetry systems in the cloud and MEC virtualized infrastructures were used to monitor resource usage, including CPU and RAM. Figure \ref{fig:videoCpuRAM} summarizes the results for \textit{high} workload scenarios and the video data type from Table \ref{tab:scenarios}. This scenario involves the most challenging data type, as indicated by the \textit{Data Access Delay} and \textit{Data Delivery Delay} results in Tables \ref{tab:resultsCloud} and \ref{tab:resultsMEC}. In the cloud infrastructure, CPU usage remains below 15\%, with RAM usage under 50\%. In contrast, the MEC infrastructure shows even more idle resources, with CPU usage below 5\% and memory usage under 15\%. It is noteworthy that resource usage remained stable throughout the tests. Additionally, in this scenario, tests conducted to verify the efficiency of policies designed to reduce overhead when different consumers perform common data queries showed negligible overhead on CPU (0.1\%) and RAM (0.08\%) in the MEC, and a minimal impact on CPU (4.5\%) and RAM (0.06\%) usage in the cloud when forking individual flows to 100 concurrent consumers.

\begin{table*}[!t]
\caption{Results with Data Consumers connected to the cloud data platform.}
\label{tab:resultsCloud}
\centering
\resizebox{\linewidth}{!}{%
\begin{tblr}{
  cells = {c},
  cell{1}{1} = {r=2}{},
  cell{1}{2} = {r=2}{},
  cell{1}{3} = {c=6}{},
  cell{3}{1} = {r=3}{},
  cell{6}{1} = {r=3}{},
  cell{9}{1} = {r=3}{},
  cell{12}{1} = {r=3}{},
  cell{15}{1} = {r=3}{},
  cell{18}{1} = {r=3}{},
  vlines,
  hline{1,3,6,9,12,15,18,21} = {-}{},
  hline{2} = {3-8}{},
  hline{4-5,7-8,10-11,13-14,16-17,19-20} = {2-8}{},
}
\textbf{Metric}                                             & \textbf{Workload} & \textbf{Data type and OBU connection} &                   &                    &                    &                    &                   \\
                                                            &                   & \textbf{C-ITS ETH}                    & \textbf{C-ITS 5G} & \textbf{Image ETH} & \textbf{Image 5G}  & \textbf{Video ETH} & \textbf{Video 5G} \\
{\textbf{Data }\\\textbf{ Access }\\\textbf{ Delay (ms)}}   & \textbf{Low}      & 956.1 (802.29)                        & 1352.8 (2143.36)  & 1027.71 (239.34)   & 1908.9 (9280.89)   & 860.8 (564.8)      & 934.8 (3021.5)    \\
                                                            & \textbf{Medium}   & 945 (1355.2)                          & 1313.7 (1266.61)  & 1049.28 (898.49)   & 1861.9 (115573.69) & 856.8 (1616.5)     & 935.6 (5650.9)    \\
                                                            & \textbf{High}     & 969.8 (2394.16)                       & 1379.1 (9494.49)  & 1085.71 (1232.77)  & 1433 (2616.8)      & 881.9 (8935.9)     & 1229.9 (96020.7)  \\
{\textbf{Data }\\\textbf{ Delivery }\\\textbf{ Delay (ms)}} & \textbf{Low}      & 37.99 (26.9)                          & 76.56 (187.35)    & 177.41 (73.41)     & 419.12 (45970.81)  & 691.7 (90.6)       & 698.9 (8801.5)    \\
                                                            & \textbf{Medium}   & 36.31 (3.98)                          & 77.12 (212.21)    & 177.13 (89.25)     & 418.25 (45913.5)   & 695.5 (100.6)      & 794.6 (875.9)     \\
                                                            & \textbf{High}     & 38.29 (10.72)                         & 81.46 (265.78)    & 196.97 (343.64)    & 440.08 (43137.44)  & 696.4 (88.2)       & 821.6 (13430.6)   \\
{\textbf{MEC }\\\textbf{ CPU }\\\textbf{ usage (\%)}}       & \textbf{Low}      & 2.26 (0.12)                           & 2.12 (0.08)       & 1.63 (0.06)        & 3.39 (0.08)        & 1.89 (0.4)         & 3.48 (0.11)       \\
                                                            & \textbf{Medium}   & 2.71 (0.12)                           & 2.52 (0.24)       & 1.73 (0.08)        & 3.42 (0.09)        & 1.98 (0.6)         & 3.52 (0.09)       \\
                                                            & \textbf{High}     & 2.86 (0.07)                           & 2.81 (0.13)       & 1.72 (0.51)        & 1.91 (0.34)        & 2.01 (0.39)        & 2.75 (0.88)       \\
{\textbf{MEC }\\\textbf{ Memory }\\\textbf{ usage (\%)}}    & \textbf{Low}      & 13.39 (0.002)                         & 13.41 (0.001)     & 13.27 (0.0005)     & 16.38 (0.03)       & 11.67 (0.02)       & 16.76 (0.05)      \\
                                                            & \textbf{Medium}   & 13.42 (0.001)                         & 13.44 (0.001)     & 13.27 (0.0008)     & 16.35 (0.05)       & 11.84 (0.02)       & 16.76 (0.01)      \\
                                                            & \textbf{High}     & 13.13 (0.005)                         & 13.11 (0.002)     & 12.96 (0.005)      & 12.49 (0.02)       & 12.23 (0.06)       & 14.35 (4.31)      \\
{\textbf{Cloud }\\\textbf{ CPU }\\\textbf{ usage (\%)}}     & \textbf{Low}      & 13.92 (0.87)                          & 12.08 (0.37)      & 11.001 (0.18)      & 16.1 (5.93)        & 11.70 (0.49)       & 17.61 (2.34)      \\
                                                            & \textbf{Medium}   & 14.94 (1.25)                          & 12.81 (0.26)      & 11.33 (0.34)       & 16.6 (3.45)        & 12.35 (0.54)       & 17.35 (0.74)      \\
                                                            & \textbf{High}     & 14.63 (2.52)                          & 13.2 (1.74)       & 11.95 (0.31)       & 12.42 (1.89)       & 11.88 (0.54)       & 15.01 (11.99)     \\
{\textbf{Cloud }\\\textbf{ Memory }\\\textbf{ usage (\%)}}  & \textbf{Low}      & 43.1 (0.0006)                         & 43.11 (0.0007)    & 43.1 (0.002)       & 44.46 (0.04)       & 45.31 (0.02)       & 44.21 (0.002)     \\
                                                            & \textbf{Medium}   & 43.11 (0.0003)                        & 43.11 (0.0005)    & 43.11 (0.0007)     & 44.43 (0.03)       & 46.24 (0.02)       & 44.23 (0.002)     \\
                                                            & \textbf{High}     & 43.05 (0.001)                         & 43.02 (0.001)     & 43.07 (0.0007)     & 39.78 (0.008)      & 46.83 (0.02)       & 40.53 (10.92)     
\end{tblr}
}
\end{table*}

\begin{table*}[!t]
\caption{Results with Data Consumers hosted by the MEC platform.}
\label{tab:resultsMEC}
\centering
\resizebox{\linewidth}{!}{%
\begin{tblr}{
  cells = {c},
  cell{1}{1} = {r=2}{},
  cell{1}{2} = {r=2}{},
  cell{1}{3} = {c=6}{},
  cell{3}{1} = {r=3}{},
  cell{6}{1} = {r=3}{},
  cell{9}{1} = {r=3}{},
  cell{12}{1} = {r=3}{},
  vlines,
  hline{1,3,6,9,12,15} = {-}{},
  hline{2} = {3-8}{},
  hline{4-5,7-8,10-11,13-14} = {2-8}{},
}
\textbf{Metric}                                             & \textbf{Workload} & \textbf{Data type and OBU connection} &                   &                    &                   &                    &                   \\
                                                            &                   & \textbf{C-ITS ETH}                    & \textbf{C-ITS 5G} & \textbf{Image ETH} & \textbf{Image 5G} & \textbf{Video ETH} & \textbf{Video 5G} \\
{\textbf{Data }\\\textbf{ Access}\\\textbf{ Delay (ms)}}    & \textbf{Low}      & 1.16 (0.05)                           & 34.09 (69.52)     & 6.58 (6.74)        & 210.42 (485.35)   & 574.6 (35816.6)    & 793.4 (157289.8)  \\
                                                            & \textbf{Medium}   & 1.06 (0.099)                          & 36.63 (36.24)     & 5.15 (5.02)        & 207.39 (186.29)   & 623.9 (50410.8)    & 953.0 (141096.4)  \\
                                                            & \textbf{High}     & 0.98 (0.14)                           & 36.98 (38.95)     & 6.28 (8.96)        & 208.26 (69.28)    & 750.2 (32743.0)    & 1183.4 (258541.1) \\
{\textbf{Data }\\\textbf{ Delivery }\\\textbf{ Delay (ms)}} & \textbf{Low}      & 1.057 (0.09)                          & 34.23 (42.67)     & 4.27 (98.15)       & 154.63 (312.45)   & 496.6 (82.4)       & 544.3 (2290.6)    \\
                                                            & \textbf{Medium}   & 0.97 (0.06)                           & 33.92 (44.25)     & 4.24 (104.73)      & 155.18 (304.79)   & 506.2 (2636.4)     & 546.0 (1668.6)    \\
                                                            & \textbf{High}     & 0.89 (0.09)                           & 33.49 (47.62)     & 3.92 (1.96)        & 154.86 (351.53)   & 505.3 (1889.6)     & 548.0 (2272.8)    \\
{\textbf{MEC}\\\textbf{ CPU}\\\textbf{ usage (\%)}}         & \textbf{Low}      & 1.99 (0.74)                           & 2.32 (1.11)       & 2.07 (1.22)        & 2.41 (1.33)       &         1.43 (0.18)           &        1.45 (0.19)           \\
                                                            & \textbf{Medium}   & 2.02 (0.87)                           & 1.87 (0.72)       & 2.08 (0.8)         & 2.17 (1.01)       &   1.67 (0.79)                 &  1.56 (0.61)                 \\
                                                            & \textbf{High}     & 2.62 (1.24)                           & 2.44 (1.43)       & 2.22 (1.16)        & 1.96 (1.09)       &   1.71 (0.18)                 &    1.62 (0.18)               \\
{\textbf{MEC }\\\textbf{ Memory}\\\textbf{ usage (\%)}}     & \textbf{Low}      & 12.57 (0.03)                          & 12.61 (0.04)      & 12.6 (0.03)        & 12.59 (0.03)      &         16.47 (0.01)           &       16.47 (0.1)            \\
                                                            & \textbf{Medium}   & 12.57 (0.02)                          & 12.51 (0.02)      & 12.61 (0.03)       & 12.57 (0.04)      &   16.46 (0.01)                 &    16.47 (0.01)               \\
                                                            & \textbf{High}     & 12.67 (0.35)                          & 12.62 (0.03)      & 12.64 (0.04)       & 12.53 (0.03)      &   16.47 (0.01)                 &    16.50 (0.05)               
\end{tblr}
}
\end{table*}

\section{Conclusion}
\label{conclusions}
Given the diverse traffic, road, and weather conditions, continuous expansion of datasets is vital for constant updates of on-board driving assistance systems’ ML models. This enhances system reliability and reduces time to market. This paper’s key contribution is the proposal of an innovative edge-cloud data platform that connects car data producers with various services, addressing critical challenges such as data sovereignty, governance, interoperability, and privacy.

The key contributions of this work are: a novel data streaming architecture that meets data space requirements by enabling efficient data filtering and preparation closer to the data source, optimizing cloud resource usage, and eliminating the need for dedicated vehicle fleets for dataset generation; a performance analysis of the platform, which translates data space requirements into practical platform features that guide system design to prevent bottlenecks, ensuring scalable data processing through the distribution of data functions across edge and cloud architectures, leading to faster data delivery and real-time performance monitoring; and an exploration of dynamic workload management, examining the effects of different data workloads—such as concurrency, data complexity, and cloud versus edge processing—on latency, scalability, and system efficiency.

The study presents a thorough performance assessment, showing minimal CPU (4\%) and RAM (2\%) fluctuation under varying workloads. The use of MEC (Multi-Access Edge Computing) for local data processing significantly reduces latency (e.g., 40ms saved for C-ITS messages), validating its suitability for time-sensitive applications. While higher workloads introduce slight delays in data access, no major bottlenecks were observed, confirming the platform's efficiency. Additionally, 5G SA connectivity, combined with MEC hosting, ensures low-latency data handling, thus supporting the real-time demands of CCAM applications.

\section*{Acknowledgements}
This work is part of the 5GMETA project, funded from the European Union's Horizon 2020 research and innovation programme under grant agreement No 957360.

\bibliographystyle{IEEEtran}
\bibliography{main}











\vfill\null

\section*{Biography Section}

\vskip -2\baselineskip plus -1fil
\begin{IEEEbiography}[{\includegraphics[width=1in,height=1.25in,clip,keepaspectratio]{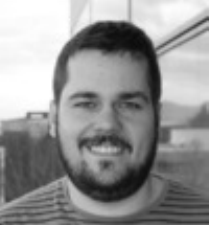}}]{Felipe Mogoll\'on} is with the Department of Digital Media, Vicomtech, Spain. He received his Telecommunication engineering degree in 2006 from Universidad de Cantabria, Spain. 
Former developer at Zitralia Security Solutions (October 2006 - June 2008). Currently, he is working at Vicomtech in multimedia services and 5G infrastructures projects.
\end{IEEEbiography}

\vskip -2\baselineskip plus -1fil

\begin{IEEEbiography}[{\includegraphics[width=1in,height=1.25in,clip,keepaspectratio]{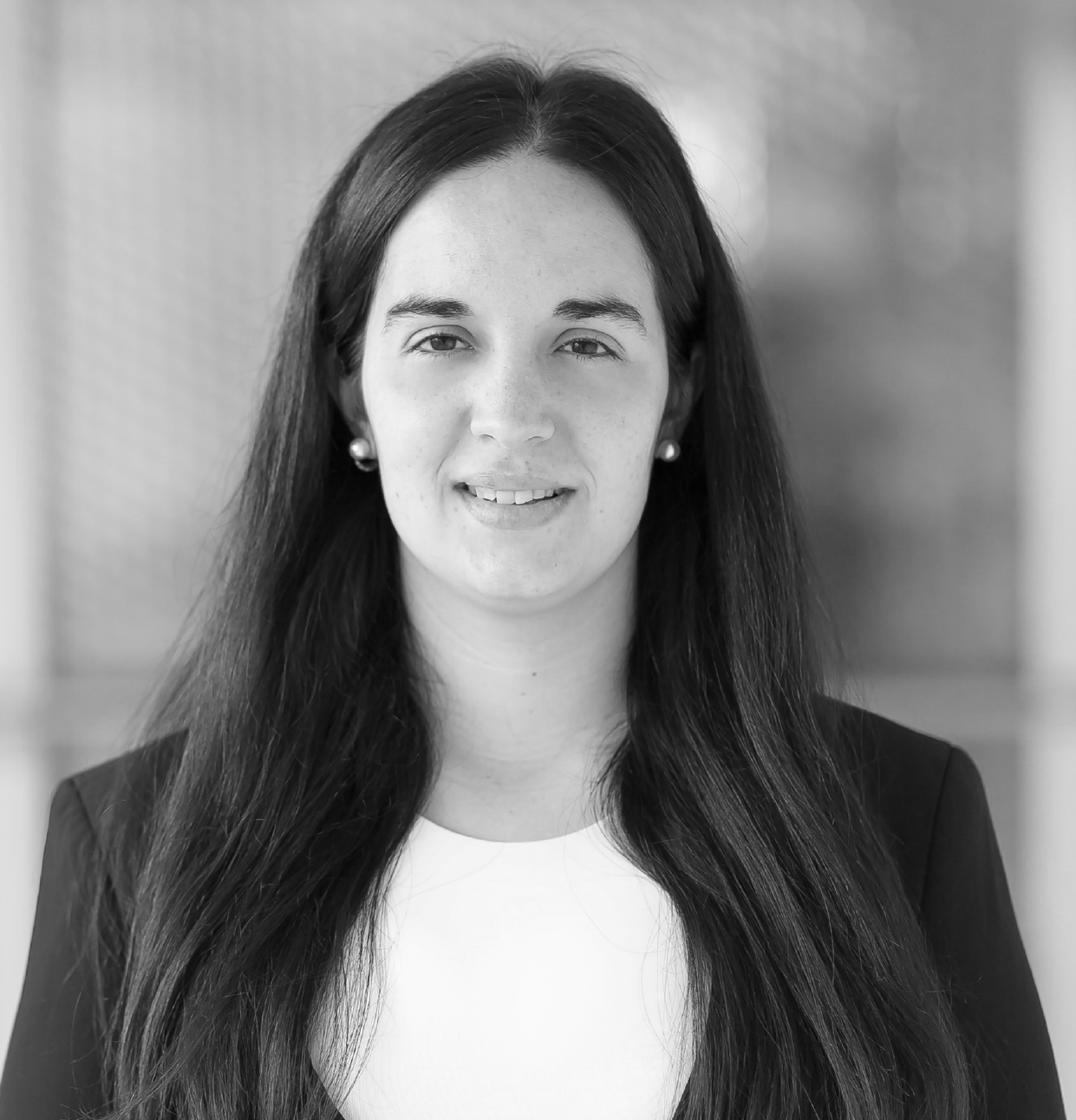}}]{Zaloa Fern\'{a}ndez} received a BSc degree in telecommunications systems and an MSc degree in embedded systems from the University of Mondragon, Spain, in 2014 and 2016, respectively, and a PhD degree from the University of Mondragon in 2019. She is a Research Associate in Vicomtech involved in projects dealing with 5G communications, virtualization paradigm, handover, network slicing and software-defined radio.
\end{IEEEbiography}

\vskip -2\baselineskip plus -1fil

\begin{IEEEbiography}[{\includegraphics[width=1in,height=1.25in,clip,keepaspectratio]{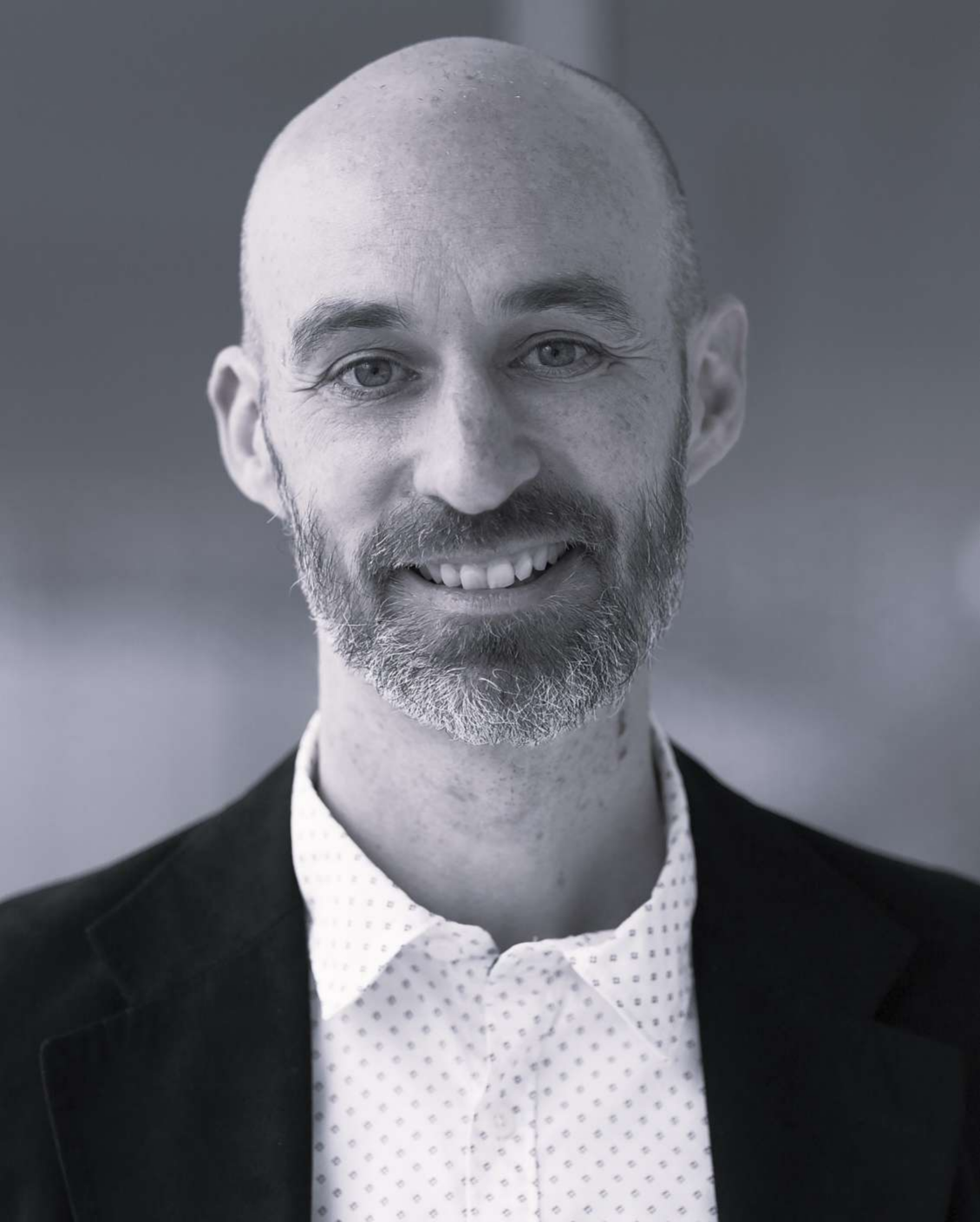}}]{\'{A}ngel Mart\'{i}n} is with the Department of Digital Media, Vicomtech. He received his PhD degree (2018) from UPV/EHU and his engineering degree (2003) from University Carlos III. He worked in media streaming and encoding research at Prodys (2003-2005) and Telefonica (2005-2008). Then, he worked in the field of ubiquitous and pervasive computing at Innovalia (2008-2010). Currently, he is at Vicomtech, working in multimedia services and 5G infrastructures projects.
\end{IEEEbiography}

\vskip -2\baselineskip plus -1fil

\begin{IEEEbiography}[{\includegraphics[width=1in,height=1.25in,clip,keepaspectratio]{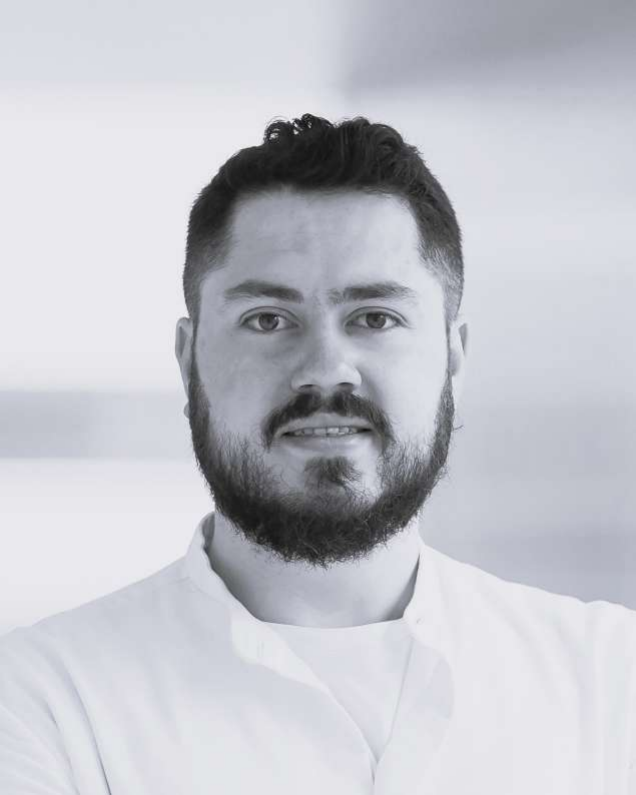}}]{Juan Diego Ortega} received the M.S. degree in mechanical engineering from the University of Navarra (UNAV), Spain, in 2013. He works as researcher on computer vision, machine learning and data processing at the Connected and Cooperative Situation Awareness Systems department at Vicomtech (Spain). His research interests, currently developing his PhD thesis, include the use of computer vision algorithms and data processing for driver monitoring systems and highly automated vehicles.
\end{IEEEbiography}

\vskip -2\baselineskip plus -1fil

\begin{IEEEbiography}[{\includegraphics[width=1in,height=1.25in,clip,keepaspectratio]{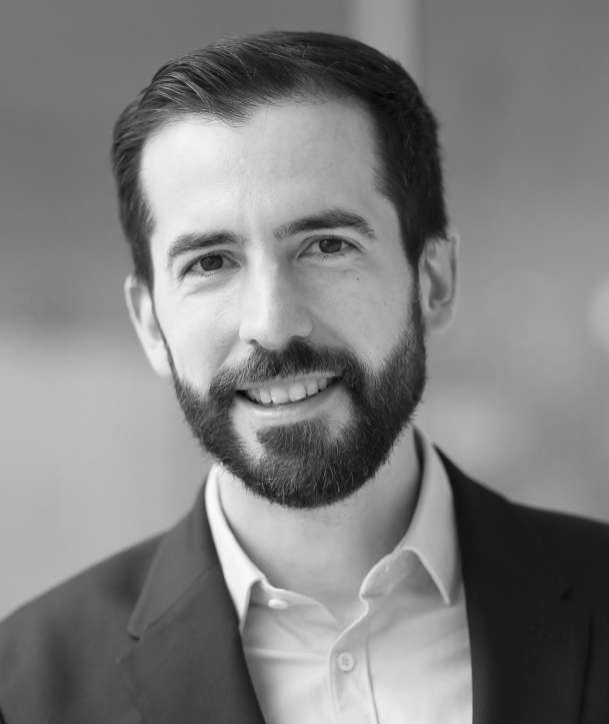}}]{Gorka V\'{e}lez} received the M.Sc. degree in Electronic Engineering from the University of Mondragon (Spain) in 2007, and the Ph.D. degree from the University of Navarra (Spain) in 2012. He leads the Connected and Cooperative Situation Awareness research line at the Intelligent Transportation Systems (ITS) and Engineering department of Vicomtech. He is the technical coordinator of the H2020 project 5GMETA. He is also involved in several other CCAM projects.
\end{IEEEbiography}

\vfill

\end{document}